\documentclass[a4paper,11pt]{article}

\usepackage{jcappub}
\usepackage{amsmath,amssymb}	
\usepackage{mathtools}
\usepackage{slashed}
\usepackage{xspace}
\usepackage{comment}
\usepackage{graphicx}
\usepackage{float}
\usepackage{siunitx}
\usepackage{hyperref}
\usepackage{aas_macros}



\title{A generic $\omega_b$ tension in early-time solutions to the Hubble tension}
\author{Cara Giovanetti}
\emailAdd{cgiovanetti@lbl.gov}
\affiliation{Theory Group, Lawrence Berkeley National Laboratory, Berkeley, CA 94720, USA}
\affiliation{Leinweber Institute for Theoretical Physics, University of California, Berkeley, CA 94720, USA}

\date{\today}

\abstract{
I show that early-time (pre-recombination) solutions to the Hubble tension are generically expected to increase the preferred baryon density $\omega_b$.  This puts these models in tension with Big Bang Nucleosynthesis (BBN), as measurements of primordial deuterium constrain $\omega_b$ at percent level.  I show that existing analyses are in tension with the BBN determination of $\omega_b$, and that including a likelihood component for primordial deuterium deters two representative models from recovering a high $H_0$.
}
\begin{document}
\maketitle
\flushbottom

\section{Introduction}
The Hubble tension remains a persistent challenge to the consistency of Lambda Cold Dark Matter ($\Lambda$CDM) cosmology.  The Cosmic Microwave Background (CMB), which is sensitive to $H_0$ through its impact on the comoving distance to recombination, tends to prefer a smaller $H_0$ than that preferred by late-time measurements of $H_0$ using the cosmic distance ladder and Type Ia supernovae.  Indeed, the determination of $H_0$ by the Planck CMB satellite~\cite{Planck2018} is $\sim5\sigma$ below the value preferred by the SH0ES cosmic distance ladder survey~\cite{Riess_2021,Riess_2022}.  

Early-time solutions to the Hubble tension argue this discrepancy can be traced to dynamics and degrees of freedom~\cite{Poulin_2019,Aloni_2022,Joseph_2023,Buen_Abad_2023}, classical fields~\cite{Jedamzik_2020}, or other effects~\cite{Hart_2017,Hart_2020} missing from $\Lambda$CDM.  Detailed numerical analyses of models like these show that the CMB can be made to prefer a large $H_0$, in better agreement with the determination of $H_0$ from supernovae alone.  These models have been tested against multiple CMB, Baryon Acoustic Oscillation (BAO), galaxy clustering, and supernova datasets and catalogs (including but not limited to Planck, ACT~\cite{Louis_2025}, SPT~\cite{camphuis2025spt3gd1cmbtemperature}, BOSS~\cite{Alam_2017}, DES~\cite{Abbott_2022}, DESI~\cite{Abdul_Karim_2025}, SH0ES, and Pantheon~\cite{Scolnic_2018,Brout_2022}), and often provide better fits to these combined datasets than $\Lambda$CDM. 

Many such models also claim consistency with constraints from Big Bang Nucleosynthesis (BBN).  BBN is sensitive to conditions in the early universe that affect the expansion rate, and so the common wisdom is that the effects of early-time solutions to the Hubble tension must not manifest until after BBN.  Some analyses have put forth explicit mechanisms to ensure these models do not meaningfully change the expansion history during BBN~\cite{Aloni_2023,allali2024}.

The models discussed above do not directly influence the baryon density, $\Omega_bh^2$ (hereafter $\omega_b$), meaning they do not, e.g., inject additional baryons or photons into the Standard Model plasma.  However, numerical analyses of these models report best-fit $\omega_b$ values much larger than the value preferred by BBN in $\Lambda$CDM (e.g., refs.~\cite{Poulin_2019,Jedamzik_2020,Aloni_2022,Buen_Abad_2023,Joseph_2023,smith2023currentsmallscalecmbconstraints,poulin2024implicationscosmiccalibrationtension,Saravanan_2025,garny2026darkacousticoscillationshubble}).  BBN provides a stringent constraint on $\omega_b$; the primordial deuterium abundance is measured to percent-level precision~\cite{Cooke_2018}, and provides sensitivity to the value of the baryon density at a similar level of precision due to the strong scaling of the deuterium yield per hydrogen $\textrm{D/H}\sim\omega_b^{-1.65}$~\cite{Pitrou_2018}.  This sensitivity is owed to efficient processing of deuterium into heavier nuclei when baryons are more abundant relative to photons.  While noted in some contexts~\cite{Vagnozzi_2021,Seto_2021,Seto_2023,takahashi2023impactbigbangnucleosynthesis,Pedrotti_2025,gonzalezfuentes2026exploringinterplaylatetimedynamical}, this constraint from BBN, and indeed this preference for large $\omega_b$, has been overlooked in many analyses of early-time solutions to the Hubble tension involving CMB data.\footnote{The role of BBN in determining $H_0$ in BAO+BBN analysis---independent of the CMB---has been explored extensively (see, e.g., refs~\cite{Addison_2018,Cuceu_2019,Schoneberg_2019}).  The baryon density enters only weakly as a calibration of the sound horizon in these analyses, though this combination still prefers a low $H_0$.}

In this work, I show early-time solutions to the Hubble tension generically prefer large $\omega_b$.  I show that this is in tension with the primordial deuterium abundance, and that considering a likelihood component for BBN in analyses with CMB data leads to a lower $H_0$ than preferred in analyses without BBN.  Below I construct a parametric argument for why the baryon density tends to increase with $H_0$ in the CMB\@.  I perform a simple analysis of two representative models, and find including this likelihood component prevents concordance with SH0ES\@.  I therefore show consistency with BBN is required to rectify the Hubble tension.

\section{Early-time solutions to the Hubble tension prefer large $\omega_b$}
In this section I argue why one should expect large $\omega_b$ in cosmologies where the CMB prefers large $H_0$.  

There are four independent angular scales measured by the CMB~\cite{Hu_Sugiyama_Silk_Nature}; the three most relevant for determining the scaling of $H_0$ with $\omega_b$ are $\ell_A$ (the extent of the sound horizon at decoupling), $\ell_{\rm{eq}}$ (the extent of the particle horizon at matter-radiation equality), and $\ell_d$ (the damping scale).  A discussion of the qualitative features of the CMB that each scale determines is available in ref.~\cite{Hu_Dodelson_2002}.  A complementary discussion of the power of these angular scales in shaping, categorizing, and assessing solutions to the Hubble tension appears in ref.~\cite{Knox_2020}.  Here, I focus explicitly on the role of the baryon density in fixing these angular scales while modifying $H_0$.  

More concretely, the angular scales are
\begin{align}
    \ell_A&=\frac{\pi D_A(z_*)}{r_s(z_*)}\nonumber\\
    \ell_{\rm{eq}}&=k_{\rm{eq}}D_A(z_*)\nonumber\\
    \ell_d&=k_dD_A(z_*).\label{eq:angular_scales}
\end{align}
$D_A(z_*)$ is the angular diameter distance at the redshift of decoupling $z_*$, given by 
\begin{equation*}
    D_A(z_*)=\frac{1}{(1+z_*)}\int_0^{z_*} \frac{dz}{H(z)},
\end{equation*}
where $H(z)$ is the Hubble parameter.  $r_s(z_*)$ is the sound horizon at decoupling, and can be written
\begin{equation}
    r_s(z_*)=\int_{z_*}^\infty \frac{c_s(z)}{H(z)}dz,\label{eq:rs}
\end{equation}
with $c_s(z)$ the fluid sound speed as a function of redshift.  $k_{\rm{eq}}$ is the wavenumber corresponding to the horizon size at matter-radiation equality.  $k_d=\pi/r_d$, where $r_d$ is the root mean squared diffusion distance and can be estimated analytically by~\cite{Weinberg_1972,Kaiser_1983, Bond_1996,Hu_1996_a}
\begin{equation}
    r_d^2(z_*)=\pi^2\int_0^{\eta_*}\frac{d\eta}{a\sigma_T n_e }\frac{R^2+\frac{16}{15}(1+R)}{6(1+R)^2}.\label{eq:diff_distance}
\end{equation}
$\sigma_T$ is the cross section for Thomson scattering and $R$ is the baryon drag term $3\rho_b/4\rho_{\gamma}$ for baryon energy density $\rho_b$ and photon energy density $\rho_{\gamma}$.  $n_e=x_en_H\sim x_e(1-\textrm{Y}_\textrm{P})\omega_b$ for free electron fraction $x_e$, hydrogen number density $n_H$, and helium-4 mass fraction $\textrm{Y}_\textrm{P}$.  $d\eta=dt/a=da/(a^2H)$ for scale factor $a$ and time $t$, and $\eta_*$ is the horizon size at decoupling.

$r_s(z_*)$ depends strongly on $H$, and so adjustments to $H_0$ to solve the Hubble tension should be compensated by a corresponding shift in $D_A$ to preserve $\ell_A$~\cite{Hou_2013}.  However, as is made clear by Eq.~\eqref{eq:angular_scales}, doing so shifts all angular scales of the CMB, and so either the variation of $D_A$ must be tempered, or other parameters must shift in response.

In practice, consistency with data is maintained by some combination of these two options.  I will assume a flat universe and ignore late-time effects from dark energy.  Then $k_{\rm{eq}}$ is given by
\begin{equation*}
    k_{\rm{eq}}=a_{\rm{eq}}H(a_{\rm{eq}})\sim\frac{\Omega_mh^2}{\sqrt{\omega_r}},
\end{equation*}
where $\omega_r=\Omega_rh^2$ is the energy density in radiation and $h$ is the dimensionless equivalent of the Hubble constant.  I leave the matter density explicit as $\Omega_mh^2$ so the $h$ dependence can be isolated below.  The $D_A$ integral is largely dominated by the $\Omega_m$ contribution to $H(z)$, so I estimate 
\begin{equation*}
    D_A\sim \frac{1}{h\sqrt{\Omega_m}}.
\end{equation*}
Assuming fixed $T_{\rm{CMB}}$ and $N_{\rm{eff}}$ (e.g., fixed $\omega_r$), 
\begin{equation*}
    \frac{\Delta\ell_{\rm{eq}}}{\ell_{\rm{eq}}}\sim\frac{\Delta h}{h} +0.5\frac{\Delta \Omega_m}{\Omega_m}.
\end{equation*}

The scalings of $\ell_A$ and $\ell_d$ are more difficult to estimate: $\ell_A$ because of the sensitivity to the integrated history of $H$ and the nontrivial dependence of $c_s(z)$ on $\omega_b$, and $\ell_d$ because of its dependence on the integrated history of $x_e$ through $n_e$ and the complicated function of the drag term.  One can estimate (see Appendix~\ref{app:estimates}) 
$r_s~\sim \Omega_m^{-1/4}h^{-1/2}$, and therefore
\begin{equation*}
    \frac{\Delta\ell_A}{\ell_A}\sim-0.5\frac{\Delta h}{h} -0.25\frac{\Delta \Omega_m}{\Omega_m},
\end{equation*}
though this neglects the (subdominant) scaling with $\omega_b$.

For $\ell_d$, I estimate the diffusion distance as $r_d\sim\sqrt{(a_*\sigma_Tn_e)^{-1}\eta_*}$~\cite{Hu_1996}.  The integral in Eq.~\eqref{eq:diff_distance} is dominated by redshifts close to recombination, and so it is appropriate to estimate $\eta_*\sim H_0^{-1}\Omega_m^{-1/2}$, yielding an estimate for $\ell_d$
\begin{equation*}
    \frac{\Delta\ell_d}{\ell_d}\sim0.25\frac{\Delta \omega_b}{\omega_b} - 0.5\frac{\Delta h}{h} -0.25\frac{\Delta \Omega_m}{\Omega_m},
\end{equation*}
where I have assumed Saha equilibrium dominates the scaling of $x_e$ with $\omega_b$, and have neglected the subdominant scaling of the function of $R$ in the integrand of Eq.~\eqref{eq:diff_distance} with $\omega_b$.

To obtain more precise scalings, I use the public Einstein-Boltzmann solver ABCMB~\cite{abcmb} to compute the relevant Jacobian factors using forward autodifferentiation.  In the vicinity of the reported best-fit from Planck 2018 TT,TE,EE+lowE, I find
\begin{align}
    \frac{\Delta\ell_{\rm{eq}}}{\ell_{\rm{eq}}}&\sim1.01\frac{\Delta h}{h} +0.6\frac{\Delta \Omega_m}{\Omega_m}\nonumber\\
    \frac{\Delta\ell_A}{\ell_A}&\sim0.08\frac{\Delta \omega_b}{\omega_b}  -0.48\frac{\Delta h}{h} -0.15 \frac{\Delta \Omega_m}{\Omega_m}\nonumber\\
    \frac{\Delta\ell_d}{\ell_d}&\sim0.28\frac{\Delta \omega_b}{\omega_b} - 0.44 \frac{\Delta h}{h} -0.12\frac{\Delta \Omega_m}{\Omega_m}\label{eq:scalings_eqs},
\end{align}
in general agreement with intuition and the naive scalings of Eq.~\eqref{eq:angular_scales} with these parameters.  The largest corrections arise from including the dependence of $c_s$ on $\omega_b$
 and the correction to the scaling of $D_A$ with $\Omega_m$.
 
These scalings are in good agreement with those reported in Appendix A of ref.~\cite{Hu_2001}, despite a slightly different fiducial cosmology and a different methodology employed for computing the relevant Jacobians.  If I assume pairs of scales are measured perfectly, I can solve Eqs.~\eqref{eq:scalings_eqs} for the scaling of $h$ (or equivalently $H_0$) with $\omega_b$ to find
\begin{alignat}{2}
    H_0&\sim\omega_b^{1.18}\qquad \ell_d&&,\ell_{\rm{eq}}\nonumber\\
    H_0&\sim\omega_b^{0.35}\qquad \ell_A&&,\ell_{\rm{eq}}\nonumber\\
    H_0&\sim\omega_b^{3.86}\qquad \ell_d&&,\ell_A,\label{eq:all_scalings}
\end{alignat}
giving the range of scaling exponents one might expect to find in CMB data depending on the relative precision with which different angular scales are measured.

All of the scalings in Eq.~\eqref{eq:all_scalings} are positive---that is, no matter the relative precision, one should expect positive scaling of $H_0$ with $\omega_b$ (though this does not account for differences in initial conditions; see Appendix~\ref{app:decorrelation}).  I check the scaling using Planck data numerically in Appendix~\ref{app:degeneracy}, 
finding a local degeneracy $H_0\sim\omega_b^{0.97}$ in the vicinity of the Planck mean parameter values.  This fits well within the range of expected scalings from Eq.~\eqref{eq:all_scalings}.

New physics can alter this relationship, though avenues to do so are limited.  Some of the most obvious pathways to do so are already exploited in the literature, e.g., varying the effective number of neutrinos $N_{\rm{eff}}$ to disrupt the scaling of $\ell_{\rm{eq}}$, modifying the recombination history to obtain a different scaling in $\ell_d$, or adding new dominant components to modify the integrated $H$ in $\ell_A$.  However, extreme modifications of these scalings risk producing cosmologies that are not in good agreement with data at any parameter values, and so careful, detailed intervention is required to change these scalings significantly from $\Lambda$CDM. 

The relationships in Eq.~\eqref{eq:all_scalings} are toxic to BBN.  Insisting on a larger CMB $H_0$ will also raise the CMB's preferred $\omega_b$. But the primordial deuterium abundance tightly constrains the baryon density~\cite{Pitrou_2021,Pisanti_2021,Giovanetti_2025_cosmo}, preventing concordance.

\begin{figure}
    \centering
    \includegraphics[width=.6\linewidth]{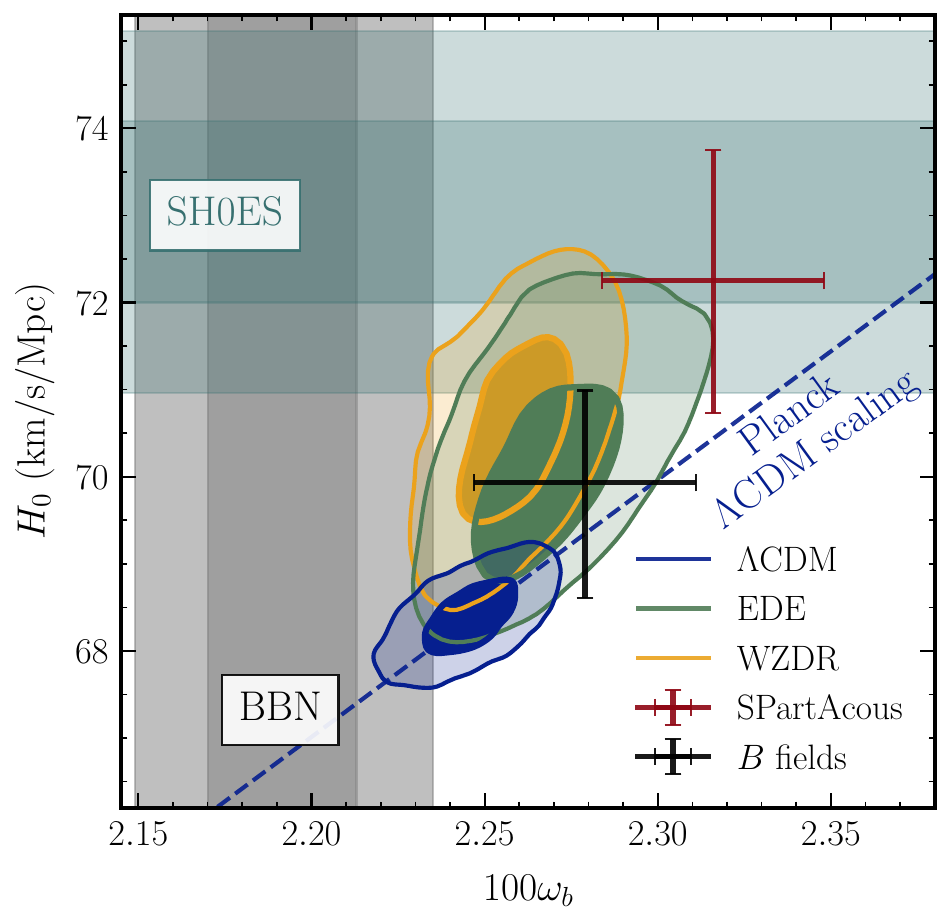}
    \caption{1 and $2\sigma$ contours in the $\omega_b-H_0$ plane for $\Lambda$CDM and early-time solutions to the Hubble tension, excluding a BBN likelihood in all analyses.  Each contour includes some combination of Planck CMB, Pantheon, SH0ES, and BAO---see Appendix~\ref{app:contour_sources} 
    for more information about each data combination.  The BBN determination of the baryon density from ref.~\cite{Giovanetti_2025_cosmo}, using the PRIMAT reaction network~\cite{Pitrou_2018}, is shown at $1$ and $2\sigma$ in shades of gray.  The SH0ES mean $\pm1$ and $2\sigma$ $H_0$~\cite{Riess_2022} are shown in light blue.  The blue dashed line is the numerically-determined scaling $H_0\sim\omega_b^{0.97}$, for $\Lambda$CDM and Planck CMB, centered on the $\Lambda$CDM contour for reference.  The $\Lambda$CDM contour includes SH0ES, hence its small size and exaggerated discrepancy with BBN (though a mild discrepancy persists even without SH0ES).  Other models shown are Early Dark Energy (EDE)~\cite{Poulin_2019}, Wess-Zumino Dark Radiation (WZDR)~\cite{Aloni_2022}, Stepped Partially Acoustic Dark Matter including three extra fermion flavors (SPartAcous)~\cite{buenabad2023b}, and primordial magnetic fields ($B$ fields)~\cite{Jedamzik_2020}.  Only the mean $\pm2\sigma$ 1D parameter values are available for SPartAcous from ref.~\cite{buenabad2023b} and for $B$ fields from ref.~\cite{Jedamzik_2025}.  Residual $H_0-\omega_b$ degeneracy remains in all cases.}
    \label{fig:H0omega_ellipses}
\end{figure}

This situation is summarized in Figure~\ref{fig:H0omega_ellipses}.  I show four models that can alleviate the Hubble tension in the $\omega_b-H_0$ plane.  Despite varying mechanisms for achieving a large $H_0$, and varying datasets used in these analyses, the trend is clear, illustrating the generic effects revealed by Eq.~\eqref{eq:all_scalings}.\footnote{A curious exception appears to occur in ref.~\cite{Hill_2022}, where, even though the positive correlation between $\omega_b$ and $H_0$ persists, some analyses nevertheless recover a smaller $\omega_b$ than $\Lambda$CDM in an Early Dark Energy cosmology.  However, this unusual relationship reflects the low-$\omega_b$-high-$n_s$ preference unique to ACT DR4~\cite{Aiola_2020}, $n_s$ the scalar spectral index.  This behavior is only realized when Planck data are excluded above $\ell=650$, and the high-$H_0$-high-$\omega_b$ trend is recovered when high-$\ell$ Planck data are included.}  

Not all early-time solutions to the Hubble tension come under pressure from this argument.  There is potential to rectify this disagreement by, perhaps counterintuitively, allowing these models to thermalize or otherwise manifest during BBN, especially if these models involve changes to $N_{\rm{eff}}$.  A large $N_{\rm{eff}}$ can increase the primordial deuterium abundance, and if $N_{\rm{eff}}$ were to increase late in BBN, there is an opportunity to fix the deuterium prediction without affecting the helium-4 prediction~\cite{Giovanetti_2025}.  However, increasing $N_{\rm{eff}}$ and increasing $\omega_b$ both have the effect of increasing Y$_{\rm{P}}$, and in light of recent precision measurements of Y$_{\rm{P}}$~\cite{aver_2026}, it is unclear whether there remains any parameter space to fix the deuterium prediction, preserve the helium-4 prediction, and still resolve the Hubble tension.  Models with additional components may instead be required; an exploration of all of these effects is left to future work.

Instead, in the next section I illustrate that models whose effects manifest after BBN provide a poor fit to CMB, BBN, BAO and supernova data when I include a full BBN likelihood.

\section{Analyses with a full BBN likelihood}As demonstrated in Figure~\ref{fig:H0omega_ellipses}, reported results are often in significant tension with the BBN determination of the baryon density.  In this section, I re-analyze two widely-studied models proposed to resolve the Hubble tension, but including a BBN likelihood in addition to CMB, BAO, and supernovae.  

Throughout, my BBN likelihood is
\begin{multline}
    -2 \log\mathcal{L}_{\rm{BBN}}= \left(\frac{\textrm{Y}_{\textrm{P}}(\omega_b ,N_{\rm{eff}},\boldsymbol{\nu}_{\rm{BBN}})-\textrm{Y}_{\textrm{P}}^{\rm{obs}}}{\sigma_{\textrm{Y}_{\textrm{P}}^{\rm{obs}}}}\right)^2 \\+ 
    \left(\frac{\textrm{D/H}(\omega_b,N_{\rm{eff}},\boldsymbol{\nu}_{\rm{BBN}})-\textrm{D/H}^{\rm{obs}}}{\sigma_{\textrm{D/H}^{\rm{obs}}}}\right)^2 \,,\label{eq:BBN_like}
\end{multline}
where
\begin{alignat*}{2}
    \textrm{D/H}^{\rm{obs}} &= 2.527\times10^{-5}\hspace{0.5cm}
    \sigma_{\textrm{D/H}^{\rm{obs}}} &&= 0.030\times10^{-5}\\
    \textrm{Y}_{\textrm{P}}^{\rm{obs}} &=0.2449  \hspace{2cm}
    \sigma_{\textrm{Y}_{\textrm{P}}^{\rm{obs}}} &&=0.004 \,.
\end{alignat*}
Despite new, precise measurements of the primordial helium-4 abundance~\cite{aver_2026}, I choose to use the older result from ref.~\cite{Aver_2015} to illustrate that the effects from including BBN are primarily driven by the inclusion of primordial deuterium in the BBN likelihood.  Given the weak scaling of $\textrm{Y}_{\textrm{P}}$ with $\omega_b$, this choice makes negligible difference in the final results.  The deuterium measurement used is from ref.~\cite{Cooke_2018}.

$\boldsymbol{\nu}_{\rm{BBN}}$ are BBN nuisance parameters---they encapsulate the uncertainties on nuclear reaction rates relevant for BBN, and on the neutron lifetime (see refs.~\cite{LINXlong, Giovanetti_2025_cosmo} for more discussion).  I therefore use the BBN code LINX~\cite{LINXlong} for BBN calculations, as it allows the user to marginalize over these parameters without hampering analysis runtime.  I use the reaction network from the BBN code PRIMAT~\cite{Pitrou_2018}.  This reaction network is known to predict a primordial deuterium abundance that is mildly discrepant with measurement~\cite{Pitrou_2021}.  However, recent work in ref.~\cite{Launders_2026} recovers this discrepancy at fixed $\omega_b$, so I therefore take this reaction network as fiducial.  This has the effect of making large $\omega_b$ even more penalized, since $\omega_b$ preferred by Planck alone in $\Lambda$CDM is already slightly too large in light of this data.  If future work pushes the $\Lambda$CDM deuterium prediction to a larger value without a substantial reduction in the prediction error, the corresponding BBN constraints on early-time solutions to the Hubble tension will weaken.

I perform two analyses, using the same data combination for each.  The first analysis is inspired by ref.~\cite{Aloni_2022}, where I test the Wess-Zumino Dark Radiation (WZDR) model proposed therein.  I use their $\mathcal{D}+$ data combination, including Planck 2018, TT,TE,EE+lowE, Planck 2018 lensing, BAO from BOSS DR12~\cite{Alam_2017}, 6dF~\cite{Beutler_2011}, and MGS~\cite{Ross_2015}, and cosmic distance ladder measurements from Pantheon~\cite{Scolnic_2018} and SH0ES~\cite{Riess_2021}, in addition to the BBN likelihood described above.  I assume new species thermalize after BBN, e.g., there is no change to $N_{\rm{eff}}$ before or during BBN.  I use ABCMB for CMB predictions, using OL\'E~\cite{gunther2025}\footnote{I modified OL\'E and its dependencies for compatibility with JAX 0.8, as is required for ABCMB.} to train an emulator on the ABCMB output and speed up the sampling procedure using ordinary, non-differentiable Markov Chain Monte Carlo (MCMC). 

In the second analysis, I use the same datasets to sample an $n=3$ Early Dark Energy (EDE) cosmology, in an analysis inspired by ref.~\cite{Poulin_2019}.  I use the CLASS~\cite{lesgourgues_2011a,lesgourgues_2011b,lesgourgues_2011c,lesgourgues_2011d} fork AxiCLASS~\cite{Smith:2019ihp,Poulin:2018dzj} for CMB in this analysis, using the same dataset combination as for the WZDR analysis.

I manage datasets and likelihoods with Cobaya~\cite{Torrado_2021}, and run four MCMC chains until the Gelman-Rubin convergence test is passed at $|R-1|<0.05$.  I run a $\Lambda$CDM analysis with the same datasets as described above, and for each cosmology ($\Lambda$CDM, WZDR, and EDE), I perform an analysis with and without the BBN component of the likelihood.  I use massless neutrinos throughout for computational efficiency---while this choice does have a non-negligible impact on the late universe through changes to $\Omega_m$, these effects are not large enough to change the conclusions of this analysis.

\begin{figure}
    \centering
    \includegraphics[width=.45\linewidth]{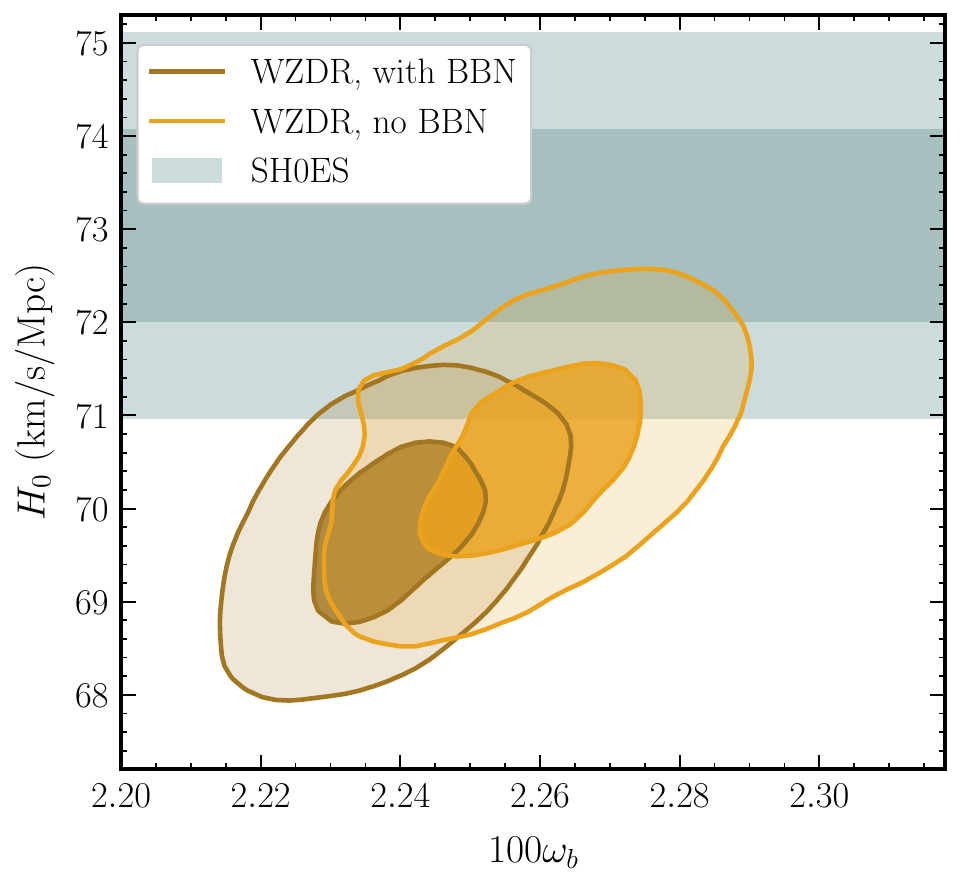}
    \includegraphics[width=.45\linewidth]{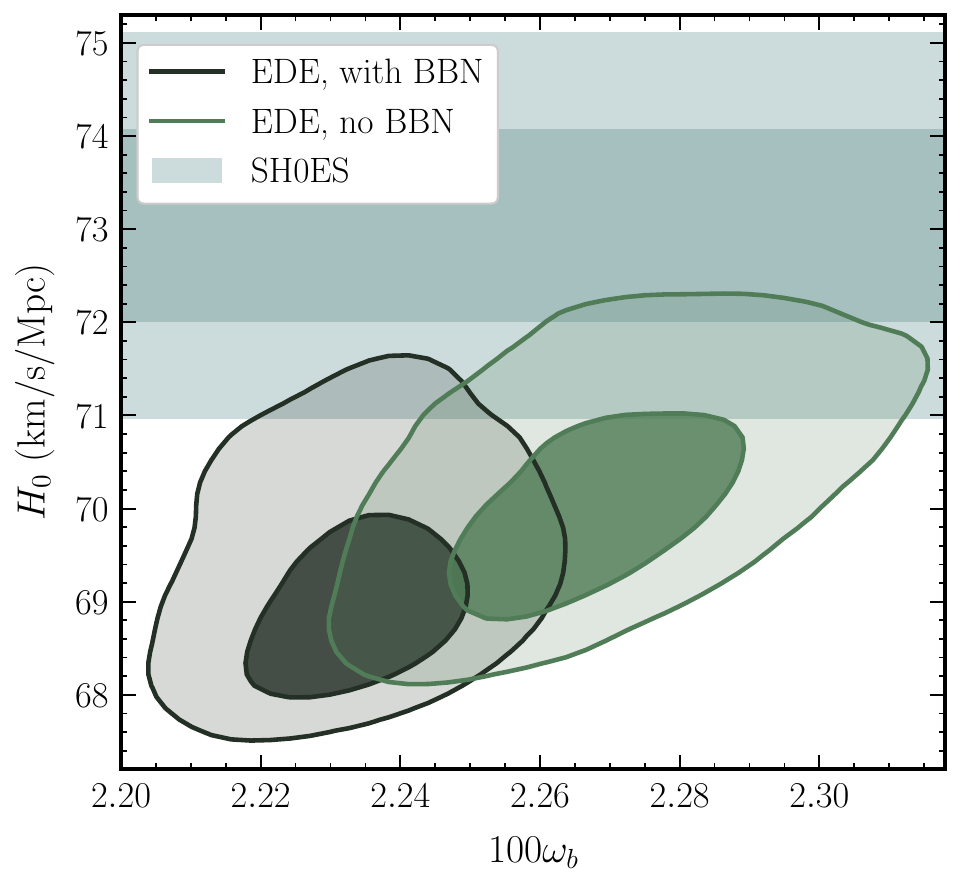}
    \caption{Results from the WZDR (left) and EDE (right) analyses described in text, with BBN (darker) and without BBN (lighter).  Results are shown at 1 and $2\sigma$.  Both analyses include a SH0ES likelihood.  The SH0ES measurement of $H_0$ is shown in light blue at $1$ and $2\sigma$.  When BBN is included, neither model can achieve as high a mean $H_0$ as without BBN.}
    \label{fig:results}
\end{figure}

Results from these analyses are summarized by Figure~\ref{fig:results}.  The effect of adding the BBN likelihood confirms expectations set by the scaling arguments: since BBN constrains $\omega_b$, the parameter degeneracy between $\omega_b$ and $H_0$ forbids these models from recovering as high $H_0$ as preferred in analyses without BBN.  Additional results from these analyses are provided in Appendix ~\ref{app:results}.

\section{Discussion}I have shown that attempts to increase the CMB-preferred value of $H_0$ will typically lead the CMB to also prefer a large $\omega_b$.  This leads early-time cosmological solutions to the Hubble tension to disagree with BBN, and additional modifications or interventions are required to recover high $H_0$.  

The explicit analyses above only considered two representative models, and did not consider effects from massive neutrinos or the combination with additional datasets.  Investigation of these effects and others is left to future work.  However, the scaling argument presented above is related to that discussed in refs.~\cite{Cyr_Racine_2022,Ge_2023}, which fixes $\omega_b$ and instead compensates high $H_0$ with Y$_{\rm{P}}$.  $\omega_b$ impacts CMB angular scales other than the damping scale, and so those arguments do not map exactly to the one considered here, but the results of this analysis are also driven in large part by the damping scale.  The conclusions, therefore, are similar: the most promising solutions to the Hubble tension are those that impact the physics of recombination (to avoid driving $\omega_b$ or Y$_{\textrm{P}}$ to tension with BBN measurement), and those that explicitly modify BBN to offset CMB degeneracies exploited to achieve large $H_0$.  Indeed, analyses of some models that modify recombination estimate~\cite{Sekiguchi_2021,Baryakhtar_2024,schoneberg2025} or find~\cite{lynch_2024} a weaker (but still positive) scaling between $\omega_b$ and $H_0$.

Other work has also put increasing pressure on late-time solutions to the Hubble tension.  Ref.~\cite{Aylor_2019} presents a clean scaling argument demonstrating why late-time solutions to the Hubble tension are generally more difficult to realize than early-time solutions (though exceptions have been found, e.g., ref.~\cite{Archidiacono_2022}).  The combination of the results from the present work and refs.~\cite{Aylor_2019,Cyr_Racine_2022,Ge_2023} therefore strongly constrains the space of possible solutions to the Hubble tension.  Concordance with all existing datasets will require awareness of these constraints.

\section*{Acknowledgments}
I am especially grateful to Neal Weiner for suggestions and feedback on this work.  I thank Martin Schmaltz and Nils Sch\"oneberg for their input on the parametric scaling argument, and Zilu Zhou for help setting up and validating WZDR in ABCMB.  I also thank Glennys Farrar, Mariangela Lisanti, Hongwan Liu, Clark Miyamoto, Vivian Poulin, Wenzer Qin, Ben Safdi, Inbar Savoray, Tristan Smith, Zach Weiner, and Zilu Zhou for helpful discussions.  I thank Miguel Escudero Abenza and Nils Sch\"oneberg for their detailed feedback on draft versions of this manuscript.  I am supported by the Office of High Energy Physics of the U.S. Department of Energy under contract DE-AC02-05CH11231. This research used resources of the National Energy Research Scientific Computing Center (NERSC), a Department of Energy User Facility (project m3166).  In addition to the packages used to run ABCMB, AxiCLASS, and Cobaya, this work makes use of the corner~\cite{Foreman-Mackey2016}, matplotlib~\cite{Hunter:2007}, NumPy~\cite{numpy:2011}, and JAX~\cite{jax2018github,deepmind2020jax} packages.


\clearpage 
\appendix
\setcounter{equation}{0}
\setcounter{figure}{0}
\setcounter{table}{0}
\renewcommand{\thefigure}{\thesection\arabic{figure}}
\renewcommand{\thetable}{\thesection\arabic{table}}
\renewcommand{\theequation}{\thesection\arabic{equation}}
\section{Estimation of scales}\label{app:estimates}
In this Appendix I detail the scaling arguments used to estimate the scaling of $\ell_A$ with $h$ and $\Omega_m$ in the main text.  The main difficulty in performing this estimate analytically is that the integral appearing in Eq.~\eqref{eq:rs} of the main body
\begin{equation}
    I=\int_0^{a_*}\frac{da}{a^2H}\label{eq:Hint}
\end{equation}
is sensitive to both the epochs of radiation and matter domination.  To estimate the scaling of this integral with cosmological parameters, I first note
\begin{equation*}
    \left(aH\right)^2\sim\omega_ra^{-2}+\omega_ma^{-1},
\end{equation*}
neglecting the very small contribution from dark energy, and now using the notation $\omega_m=\Omega_mh^2$.  Introducing the variable $y=a/a_{\rm{eq}}$ and noting $a_{\rm{eq}}=\omega_r/\omega_m$, I rewrite the above as
\begin{equation}
    \left(aH\right)^2\sim\frac{\omega_m^2}{\omega_r}\frac{1+y}{y^2},\label{eq:Hy}
\end{equation}
where no new approximations have been introduced between Eqs.~\eqref{eq:Hint} and~\eqref{eq:Hy}.  I then rewrite Eq.~\eqref{eq:Hint} as
\begin{equation*}
    I=\frac{\sqrt{\omega_r}}{\omega_m}\int_0^{y_*}\frac{dy}{\sqrt{1+y}},
\end{equation*}
where $y_*=a_*/a_{\rm{eq}}\approx3$.  Note this integral over $y$ can be evaluated analytically:
\begin{equation}
    I=2\frac{\sqrt{\omega_r}}{\omega_m}\sqrt{1+y}|_{y=0}^{y=y_*}\approx2\frac{\sqrt{\omega_r}}{\omega_m}.\label{eq:numerical_I}
\end{equation}

Let $F(y_*)=\int_0^{y_*}dy/\sqrt{1+y}$.  In order to find the scaling of $r_s$ with $\omega_m$, I need the scaling 
\begin{align}
    \frac{\partial \ln I}{\partial \ln\omega_m}&=\frac{\partial\ln\frac{\sqrt{\omega_r}}{\omega_m}}{\partial\ln\omega_m} + \frac{\partial\ln F(y_*)}{\partial \ln y_*}\frac{\partial\ln y_*}{\partial\ln\omega_m}\nonumber\\
    &=-1+\frac{y_*}{\sqrt{1+y_*}F(y_*)}\label{eq:dIdom},
\end{align}
where $y_*\sim\omega_m$ follows from the definition of $y$ and the assumption that $a_*$ depends only weakly on cosmological parameters~\cite{Hu_2001}.  

In the main text, this result is used to estimate the scaling of $r_s$ with $\omega_m$ and therefore $h$ and $\Omega_m$.  Using results from Eqs.~\eqref{eq:numerical_I} and~\eqref{eq:dIdom}, I find
\begin{equation}
    \frac{\partial \ln r_s}{\partial \ln\omega_m}\approx-1+\frac{3}{4}=-\frac14.
\end{equation}
This suggests $r_s\sim\Omega_m^{-1/4}h^{-1/2}$, as used in the main text.

\section{Numerical degeneracy analysis}\label{app:degeneracy}

The goal of this appendix is to numerically verify the parametric scaling derived in the main body.  I provide the details of the numerical analysis used to estimate the local degeneracy between $H_0$ and $\omega_b$ in $\Lambda$CDM, using Planck TT,TE,EE+lowE data, where I reported in the main body $H_0\sim\omega_b^{0.97}$.  I also include a comparison of this scaling with that of other parameters to illustrate the other available degeneracy directions to achieve a higher $H_0$.  

I follow a procedure inspired by that in ref.~\cite{Kable_2019}, rather than  full six-parameter principal component analysis.  Given that I am extrapolating these results to non-$\Lambda$CDM cosmologies, and I am primarily interested only in pairwise scalings, this proxy is sufficient as a numerical check of the intuition from the main body.   First, in Appendix~\ref{app:covariance}, I transform the Planck covariance matrix from its native parameterization (which uses $\theta_{\rm{MC}}$ rather than $H_0$) into a covariance over $H_0$ directly, using a local Jacobian evaluated with CAMB. This transformation is performed in the natural, linear parameter space. 

Second, in Appendix~\ref{app:decorrelation}, I extract the log-slope $c = d\ln H_0 / d\ln \omega_b$ along the posterior ridge.  Since $H_0 \sim \omega_b^c$ is only a linear relationship in log space, passing from the linear-space covariance to $c$ requires the approximation $\delta \ln x \approx \delta x/\bar{x}$, which only holds when the posterior is sufficiently narrow.  I validate this assumption explicitly by drawing numerical samples from the linear-space Gaussian, converting those individual samples to log space, and then computing the log-slope directly from those samples.  This procedure does not require a narrow posterior, and its agreement with the estimate obtained from transforming the linear covariance matrix directly confirms the posterior is sufficiently narrow for these estimates to be accurate.

\subsection{Covariance}\label{app:covariance}

I use the \texttt{base\_plikHM\_TTTEEE\_lowE.covmat} covariance matrix provided by Planck~\cite{Planck2018} and obtained from Cobaya~\cite{Torrado_2021}.\footnote{\url{https://github.com/CobayaSampler/planck_supp_data_and_covmats/blob/master/covmats/base_plikHM_TTTEEE_lowE.covmat}}  The reported covariances use the Cobaya $\theta_{\rm{MC}}$ parameter, and so I use CAMB~\cite{Lewis_1999,Howlett_2012} to convert these into entries in $H_0$.  This section provides the details of that transformation.

With the original parameter vector $\mathbf{x} = (\omega_b, \Omega_{\rm{CDM}}h^2, \theta_{\rm{MC}}, \tau_{\rm{reio}},\dots)$,
and covariance matrix $C_x = \langle \delta \mathbf{x}\,\delta \mathbf{x}^{\rm T} \rangle$, 
I define a new parameter vector $\mathbf{y} = (\omega_b, \Omega_{\rm{CDM}}h^2, H_0, \tau_{\rm{reio}},\dots)$, which is identical to $\mathbf{x}$ apart from the substitution of $\theta_{\rm{MC}}$ with $H_0$.  $H_0$ is explicitly a function of $\omega_b$, $\Omega_{\rm{CDM}}h^2$, and $\theta_{\rm{MC}}$.

Locally, near a fiducial point $\bar{\mathbf{x}}$ (here, the parameter means reported by Planck), this change of variables is linearized as
\begin{equation*}
\delta y_a = \sum_i J_{ai}\,\delta x_i,
\end{equation*}
with Jacobian
\begin{equation*}
J_{ai} = \left.\frac{\partial y_a}{\partial x_i}\right|_{\bar{\mathbf{x}}}.
\end{equation*}
The covariance matrix for the new vector $\mathbf{y}$ can be obtained from the first via
\begin{equation*}
C_y = J\, C_x\, J^{\rm T},
\label{eq:cov-transform}
\end{equation*}
or in components, 
\begin{equation*}
{\rm Cov}(y_a,y_b)=\sum_{ij} J_{ai}\,{\rm Cov}(x_i,x_j)\,J_{bj}.
\end{equation*}

Since all coordinates other than $\theta_{\rm{MC}}$ are unchanged, the Jacobian is the identity matrix except for the row corresponding to $H_0$.  Further, $H_0$ only has explicit dependence on a subset of the parameters in $\mathbf{x}$.  Dropping terms that are 0, I can write
\begin{equation*}
\delta H_0
=
\frac{\partial H_0}{\partial (\omega_b )}\,\delta(\omega_b)
+
\frac{\partial H_0}{\partial (\Omega_{\rm CDM} h^2)}\,\delta(\Omega_{\rm CDM} h^2)
+
\frac{\partial H_0}{\partial \theta_{\rm{MC}}}\,\delta\theta_{\rm{MC}},
\label{eq:dH0-linear}
\end{equation*}
so that the $H_0$ row of $J$ is
\begin{equation*}
J_{H_0,i}
=
\left(
\frac{\partial H_0}{\partial (\omega_b)},
\frac{\partial H_0}{\partial (\Omega_{\rm CDM} h^2)},
\frac{\partial H_0}{\partial \theta_{\rm{MC}}}
\right)
\end{equation*}
in the appropriate columns $i$, with zeros elsewhere.  Equivalently, the transformed covariance elements involving $H_0$ are
\begin{equation*}
{\rm Cov}(H_0,y_b)
=
\sum_i
\frac{\partial H_0}{\partial x_i}\,
{\rm Cov}(x_i,x_b),
\end{equation*}
and
\begin{equation*}
{\rm Var}(H_0)
=
\sum_{ij}
\frac{\partial H_0}{\partial x_i}
\frac{\partial H_0}{\partial x_j}
{\rm Cov}(x_i,x_j).
\end{equation*}

Transforming the provided covariance matrix boils down, then, to computing these derivatives; I use finite differences about the Planck means for their TT,TE,EE+lowE analysis for this purpose, computing $H_0(\omega_b,\Omega_{\rm{CDM}} h^2,\theta_{\rm{MC}})$ with CAMB.

After constructing the covariance matrix, I run CAMB again at the Planck best-fit parameters to estimate the inferred $H_0$ mean (from converting the Planck best-fit $\theta_{\rm{MC}},\omega_b,\Omega_{\rm{CDM}}h^2$ directly to $H_0$ with CAMB at its default settings) and uncertainty (using the derived covariance matrix).  While this procedure is only approximate, it recovers the distribution $H_0=67.25\pm\SI{0.61}{km/s/Mpc}$, in good agreement with the Planck reported value of $H_0=67.27\pm\SI{0.60}{km/s/Mpc}$.

\subsection{Decorrelation}\label{app:decorrelation}

With the adjusted covariance matrix in hand, I use it to perform a degeneracy analysis to determine numerically the local degeneracy of $H_0$ and other $\Lambda$CDM parameters, in the vicinity of the Planck best fit parameters.  The goal is to find the exponent $c$ for which the correlation between $H_0/p^c$ and $p$ is 0, $p$ the $\Lambda$CDM parameters apart from $H_0$.  This corresponds to the value of $c$ for which there is no residual degeneracy between $H_0$ and $p$, providing a good proxy for the local relative scalings of these parameters.  

As discussed above, I perform two analyses in this section: one which assumes a narrow posterior and transforms the covariance matrix obtained from the previous section into log space, and another that explicitly checks this assumption by sampling the linear-space covariance matrix and converting samples to log space.  I describe each procedure in detail below.

For each parameter $p \in\left(\omega_b,\Omega_{\rm{CDM}}h^2,\tau_{\rm{reio}},n_s,A_s\right)$, I truncate the adjusted covariance matrix from the previous subsection to a $2\times 2$ covariance in these parameters
\begin{equation*}
C_{\rm lin}^{(H_0,p)} =
\begin{pmatrix}
{\rm Var}(H_0) & {\rm Cov}(H_0,p) \\
{\rm Cov}(H_0,p) & {\rm Var}(p)
\end{pmatrix}.
\end{equation*}
It is easier to characterize a local scaling $H_0 \sim p^{c}$ in log space, where the relationship is linear:
\begin{equation*}
\ln H_0 \approx c \,\ln p + {\rm const.} \label{eq:linear_lower}
\end{equation*}
I therefore transform this into a covariance in log space, and I denote the covariance in the linear parameters $C_{\rm{lin}}$.

To transform the covariance matrix into log space, I assume the posterior is sufficiently narrow that I can approximate
\begin{equation}
\delta \ln H_0 \approx \frac{\delta H_0}{\bar H_0},
\qquad
\delta \ln p \approx \frac{\delta p}{\bar p},\label{eq:lin_approx}
\end{equation}
where $\bar{H}_0$ and $\bar{p}$ are the Planck TT,TE,EE+lowE means.  I check this assumption numerically below.

Then the Jacobian of the transformation to log space is
\begin{equation*}
J =
\begin{pmatrix}
1/\bar H_0 & 0 \\
0 & 1/\bar p
\end{pmatrix},
\end{equation*}
and the local log-space covariance is
\begin{equation*}
C_{\log}^{(H_0,p)} \approx J\, C_{\rm lin}^{(H_0,p)}\, J^{\rm T}.
\end{equation*}
In components,
\begin{align*}
{\rm Var}(\ln H_0) &\approx \frac{{\rm Var}(H_0)}{\bar H_0^2}, \\
{\rm Var}(\ln p) &\approx \frac{{\rm Var}(p)}{\bar p^2}, \\
{\rm Cov}(\ln H_0,\ln p) &\approx \frac{{\rm Cov}(H_0,p)}{\bar H_0 \bar p}.
\end{align*}

The decorrelation condition is easy to define in this space.  The residual
\begin{equation*}
r(c) = \ln H_0 - c \ln p
\end{equation*}
should have no dependence on $\ln p$ when $c$ captures the dependence of $\ln H_0$ on $\ln p$.  In other words, the appropriate $c$ is defined by
\begin{equation*}
{\rm Cov}\left(r(c),\ln p\right)
=
{\rm Cov}(\ln H_0,\ln p) - c\,{\rm Var}(\ln p)=0, 
\end{equation*}
which is solved by
\begin{equation}
c=\frac{{\rm Cov}(\ln H_0,\ln p)}{{\rm Var}(\ln p)}.
\label{eq:alog}
\end{equation}
This result is also the ordinary least-squares slope, with intercept, for regressing $\ln H_0$ on $\ln p$, or can also be obtained by minimizing the variance of the residual ${\rm Var}[r(c)]={\rm Var}(\ln H_0) - 2c\,{\rm Cov}(\ln H_0,\ln p) + c^2 {\rm Var}(\ln p)$ with respect to $c$.

The exponent $c$ extracted by this procedure has the local interpretation
\begin{equation}
c\approx\frac{d\ln H_0}{d\ln p}
\end{equation}
along the marginalized posterior ridge in the $(H_0,p)$ plane. It therefore provides a natural quantity to compare against the parametric scaling argument of the form $H_0 \sim p^{c}$ in the main text.

The result in Eq.~\eqref{eq:alog} is the appropriate quantity to estimate the parametric scaling between $H_0$ and a parameter $p$, as it is the most natural local derivative-like quantity that accesses the linear relationship between these parameters in log space.  However, its validity depends on the approximation in Eq.~\eqref{eq:lin_approx}.  It is not a given that this local approximation is actually accurate over the whole posterior width, as linearity of the map between the linear parameters and the log parameters may not be a good assumption.  To test the effect of the linear-to-log map explicitly, I generate 200,000 samples from
\begin{equation*}
(H_0,p) \sim \mathcal{N}\left(\bar{\mathbf p},\,C_{\rm lin}^{(H_0,p)}\right),
\end{equation*}
where $\bar{\mathbf p}$ is a vector of Planck means for all of the $\Lambda$CDM parameters.  I manually discard any samples with non-positive entries, and compute the log of each sample to recover samples in $\ln H_0$ and $\ln p$.  The resulting numerical
\begin{equation}
c_{\rm{num}}=\frac{{\rm Cov}_{\rm num}(\ln H_0,\ln p)}{{\rm Var}_{\rm num}(\ln p)}
\end{equation}
should agree with the local result above if the posterior is sufficiently narrow and close to Gaussian for the log approximation made in Eq.~\eqref{eq:lin_approx}.

Results from these procedures are tabulated in Table~\ref{tab:a_results} for all $\Lambda$CDM parameters.  The two procedures used to estimate the scaling agree to better than percent-level, and so the local, linear estimation of $c$ is sufficient despite the potential for nonlinearities.  The linear scaling falls within the range of predicted scalings from the main text.

\begin{table}[t]
    \centering
    \begin{tabular}{c|c|c}
        Parameter & $c$ & $c_{\rm{num}}$\\
        \hline
         $\omega_b$ & 0.966 & 0.969\\
         $\Omega_{\rm{CDM}}h^2$ &  -0.763 &   -0.763 \\
         $\tau_{\rm{reio}}$ & 0.0158  &   0.0153 \\
         $\ln\left(10^{10}A_s\right)$ & 0.103 & 0.105\\
         $n_s$ & 1.41 &  1.41 
    \end{tabular}
    \caption{Scaling exponents $c$ and $c_{\rm{num}}$ between the cosmological parameters in the first column and $H_0$.  The agreement between the two estimation procedures is $<1$\%, confirming that the local, linear, derivative estimation of $c$ is accurate.}
    \label{tab:a_results}
\end{table}

The scalar tilt $n_s$ did not factor into the parametric argument in the main text, and indeed I find an even stronger scaling of $H_0$ with $n_s$ in Table~\ref{tab:a_results}.  This trend is also broadly observed in analyses of models intended to solve Hubble tension (e.g., refs.~\cite{Jedamzik_2020,Aloni_2022,Hill_2022,buenabad2023b}), and indeed is observed in the analysis performed in the main body.  However, $n_s$ is constrained well enough that it cannot absorb all of the degeneracy between $H_0$ and $\omega_b$, as demonstrated empirically in the analyses performed in this manuscript.

\section{Datasets and analysis details for existing contours}\label{app:contour_sources}

Here I provide an accounting of the datasets used to obtain each result in Figure~\ref{fig:H0omega_ellipses}.  The $\Lambda$CDM, EDE, and WZDR contours are obtained in this work and use the same combination of datasets used in analyses in the main body (Planck 2018, TT,TE,EE+lowE and lensing~\cite{Planck2018}; BAO from BOSS DR12~\cite{Alam_2017}, 6dF~\cite{Beutler_2011}, and MGS~\cite{Ross_2015}; and cosmic distance ladder measurements from Pantheon~\cite{Scolnic_2018} and SH0ES~\cite{Riess_2021}).  The Stepped Partially Acoustic Dark Matter with three extra fermion flavors (SPartAcous+3) mean $\pm2\sigma$ point uses a similar combination to the other three, but also uses DES~\cite{Abbott_2022} and KiDS-1000~\cite{Heymans_2021} galaxy clustering measurements (from ref.~\cite{buenabad2023b}).  The mean $\pm2\sigma$ point for primordial magnetic fields is from ref.~\cite{Jedamzik_2025}, and uses Planck, DESI BAO~\cite{Adame_2025}, Pantheon+, and SH0ES.  Despite the use of different data combinations, the correlation between $H_0$ and $\omega_b$ is manifest.

\section{Full results from numerical analyses}\label{app:results}
\begin{figure}
    \centering
    \includegraphics[width=\linewidth]{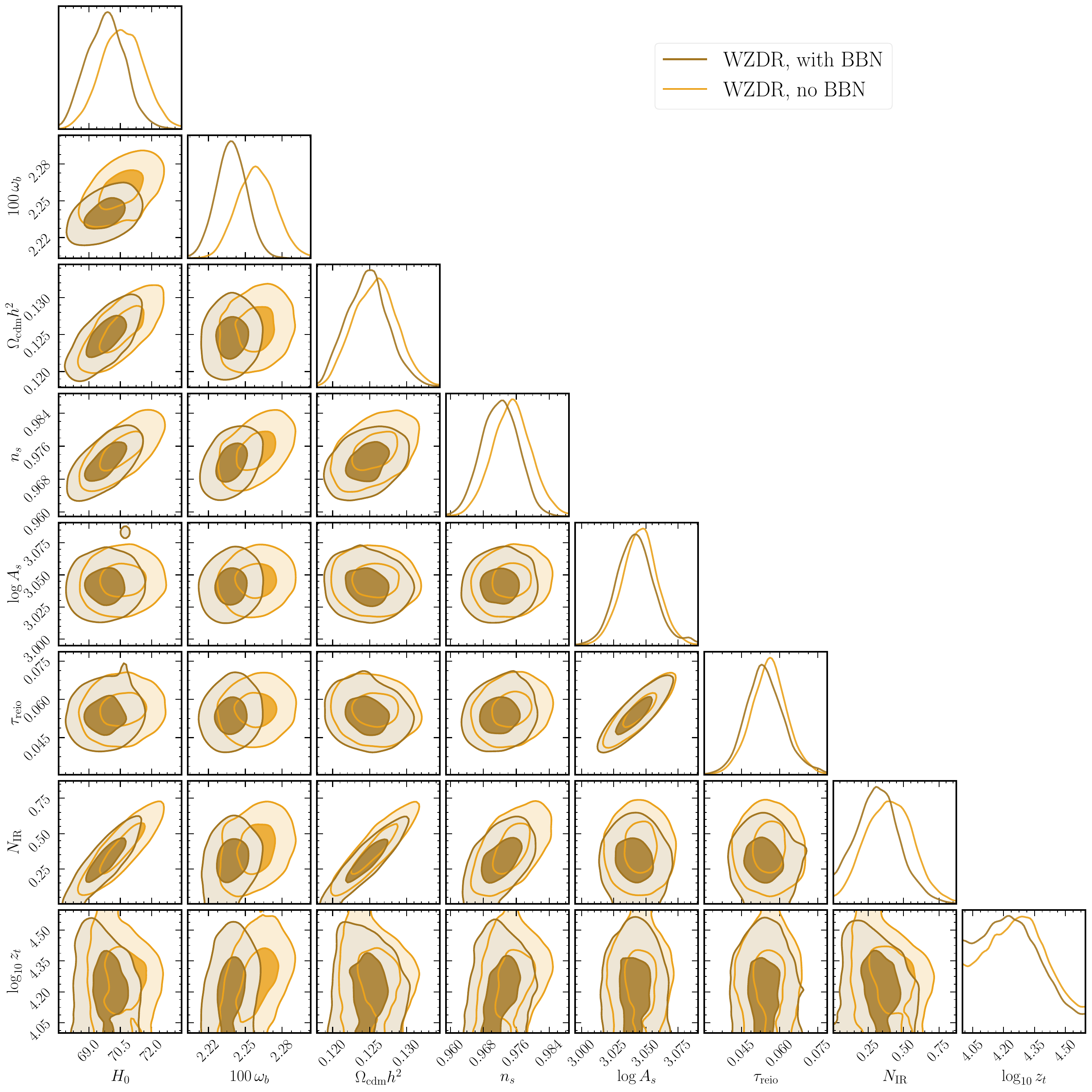}
    \caption{The effect of including a BBN likelihood in an analysis of a WZDR cosmology.  Contours are shown at 1 and $2\sigma$.  The parameters $N_{\rm{IR}}$ and $z_t$ are specific to WZDR, with the former controlling the amount of extra radiation at late times and the latter controlling the redshift of the transition in this model.  See ref.~\cite{Aloni_2022} for more information.}
    \label{fig:wzdr_corner}
\end{figure}

\begin{figure}
    \centering
    \includegraphics[width=\linewidth]{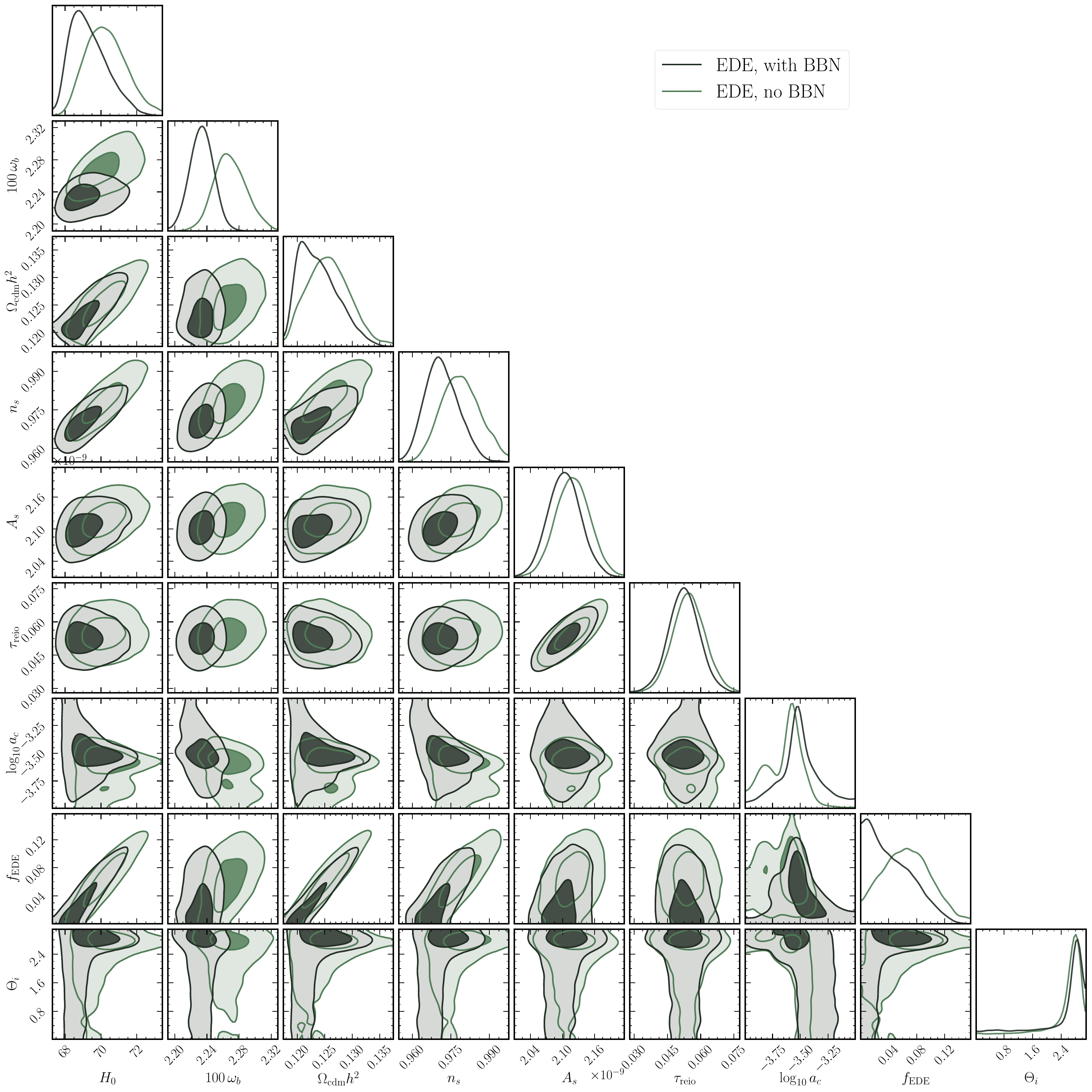}
    \caption{The effect of including a BBN likelihood in an analysis of an EDE cosmology.  Contours are shown at 1 and $2\sigma$.  $a_c$, $f_{\rm{EDE}}$, and $\Theta_i$ are parameters unique to this model, with the first controlling the scale factor at which dark energy dominates, the second controlling the fraction of dark energy added, and the third determining the initial displacement of the scalar field.  See ref.~\cite{Poulin_2019} for more information.}
    \label{fig:ede_corner}
\end{figure}

\setcounter{table}{0}
In this appendix I present full results from the two numerical analyses described in the main body.  A full corner plot for the two WZDR analyses is shown in Figure~\ref{fig:wzdr_corner}, and a full corner plot for the two EDE analyses is shown in Figure~\ref{fig:ede_corner}.  Full results are tabulated in Tables~\ref{tab:results_numbers_noBBN} (without a BBN likelihood) and~\ref{tab:results_numbers} (with a BBN likelihood).

I also use Cobaya's \texttt{--minimize} option to find the best-fit/minimum $\chi^2$ points in each analysis.  I find that both the mean and best-fit $H_0$ values across all analyses shift downward, often by roughly $1\sigma$, with the inclusion of a BBN likelihood.  The statistical preference ($\Delta$AIC) for each model does not change appreciably across analyses (the modest improvement in the EDE $\Delta$AIC can be attributed to the poorer fit of $\Lambda$CDM to both Planck high-$\ell$ TTTEEE and SH0ES in the combined analysis).  The addition of a massive neutrino has the potential to shift the means and best fits for each of these analyses---however, it is unlikely that adding a massive neutrino will significantly change the qualitative aspects of these results.

\begin{table}[htbp]
\centering
\scriptsize
\renewcommand{\arraystretch}{1.3}
\begin{tabular}{lccc}
\hline\hline
Parameter & $\Lambda$CDM & WZDR & EDE \\
\hline
$H_0$ [km/s/Mpc] & $68.40\;(68.48) \pm 0.40$ & $70.61\;(71.05) \pm 1.01$ & $70.30\;(71.92) \pm 1.10$ \\
$\omega_b$ & $0.02246\;(0.02251) \pm 0.00013$ & $0.02259\;(0.02260) \pm 0.00015$ & $0.02269\;(0.02273) \pm 0.00020$ \\
$\Omega_{\rm CDM}h^2$ & $0.11881\;(0.11866) \pm 0.00088$ & $0.1258\;(0.1284) \pm 0.0029$ & $0.1258\;(0.1300) \pm 0.0038$ \\
$\ln(10^{10}A_s)$ & $3.043\;(3.044) \pm 0.014$ & $3.045\;(3.045) \pm 0.014$ & $3.053\;(3.058) \pm 0.015$ \\
$n_s$ & $0.9673\;(0.9687) \pm 0.0036$ & $0.9750\;(0.9768) \pm 0.0048$ & $0.9789\;(0.9869) \pm 0.0074$ \\
$\tau_{\rm reio}$ & $0.0551\;(0.0546) \pm 0.0070$ & $0.0559\;(0.0549) \pm 0.0071$ & $0.0547\;(0.0549) \pm 0.0076$ \\
$N_{\rm IR}$ & --- & $0.40\;(0.52) \pm 0.16$ & --- \\
$\log_{10} z_t$ & --- & $4.27\;(4.27) \pm 0.14$ & --- \\
$f_{\rm EDE}$ & --- & --- & $0.066\;(0.109) \pm 0.034$ \\
$\log_{10} a_c$ & --- & --- & $-3.64\;(-3.58) \pm 0.16$ \\
$\Theta_i$ & --- & --- & $2.38\;(2.75) \pm 0.72$ \\
\hline
& \multicolumn{3}{c}{$\chi^2$ (best fit)} \\
\hline
Planck low-$\ell$ TT & $22.80$ & $21.65$ & $21.23$ \\
Planck low-$\ell$ EE & $396.02$ & $396.10$ & $396.04$ \\
Planck high-$\ell$ TTTEEE & $2343.53$ & $2346.39$ & $2342.05$ \\
Planck lensing & $8.76$ & $9.75$ & $9.87$ \\
6dFGS BAO & $0.01$ & $0.02$ & $0.05$ \\
SDSS DR7 MGS & $1.98$ & $2.17$ & $2.45$ \\
SDSS DR12 BAO & $3.39$ & $3.42$ & $3.69$ \\
Pantheon & $1034.74$ & $1034.74$ & $1034.90$ \\
SH0ES ($M_b$) & $7.43$ & $1.23$ & $0.33$ \\
\hline
$\chi^2_{\rm tot}$ & $3818.67$ & $3815.46$ & $3810.61$ \\
$\Delta$AIC & $0.00$ & $0.80$ & $-2.05$ \\
\hline\hline
\end{tabular}
\caption{Full results from analyses without BBN performed in this work.  All analyses use the same data combination of Planck 2018, TT,TE,EE+lowE and lensing~\cite{Planck2018}; BAO from BOSS DR12~\cite{Alam_2017}, 6dF~\cite{Beutler_2011}, and MGS~\cite{Ross_2015}; and cosmic distance ladder measurements from Pantheon~\cite{Scolnic_2018} and SH0ES~\cite{Riess_2021}.}
\label{tab:results_numbers_noBBN}
\end{table}

\begin{table}[htbp]
\centering
\scriptsize
\renewcommand{\arraystretch}{1.3}
\begin{tabular}{lccc}
\hline\hline
Parameter & $\Lambda$CDM & WZDR & EDE \\
\hline
$H_0$ [km/s/Mpc] & $68.11\;(67.95) \pm 0.38$ & $69.81\;(68.86) \pm 0.90$ & $69.33\;(70.85) \pm 1.01$ \\
$\omega_b$ & $0.02231\;(0.02225) \pm 0.00012$ & $0.02239\;(0.02234) \pm 0.00012$ & $0.02233\;(0.02241) \pm 0.00014$ \\
$\Omega_{\rm CDM}h^2$ & $0.11922\;(0.11945) \pm 0.00087$ & $0.1248\;(0.1218) \pm 0.0028$ & $0.1239\;(0.1289) \pm 0.0035$ \\
$\ln(10^{10}A_s)$ & $3.041\;(3.034) \pm 0.014$ & $3.042\;(3.055) \pm 0.015$ & $3.046\;(3.064) \pm 0.015$ \\
$n_s$ & $0.9661\;(0.9648) \pm 0.0037$ & $0.9720\;(0.9701) \pm 0.0044$ & $0.9715\;(0.9806) \pm 0.0064$ \\
$\tau_{\rm reio}$ & $0.0537\;(0.0486) \pm 0.0072$ & $0.0543\;(0.0599) \pm 0.0076$ & $0.0525\;(0.0558) \pm 0.0071$ \\
$N_{\rm IR}$ & --- & $0.32\;(0.14) \pm 0.15$ & --- \\
$\log_{10} z_t$ & --- & $4.24\;(4.32) \pm 0.14$ & --- \\
$f_{\rm EDE}$ & --- & --- & $0.044\;(0.093) \pm 0.031$ \\
$\log_{10} a_c$ & --- & --- & $-3.51\;(-3.55) \pm 0.19$ \\
$\Theta_i$ & --- & --- & $2.28\;(2.78) \pm 0.85$ \\
\hline
& \multicolumn{3}{c}{$\chi^2$ (best fit)} \\
\hline
Planck low-$\ell$ TT & $23.29$ & $22.84$ & $21.86$ \\
Planck low-$\ell$ EE & $395.72$ & $397.50$ & $396.35$ \\
Planck high-$\ell$ TTTEEE & $2346.63$ & $2344.36$ & $2341.19$ \\
Planck lensing & $8.89$ & $9.20$ & $10.34$ \\
6dFGS BAO & $0.00$ & $0.00$ & $0.00$ \\
SDSS DR7 MGS & $1.53$ & $1.73$ & $1.92$ \\
SDSS DR12 BAO & $3.74$ & $3.51$ & $3.42$ \\
Pantheon & $1035.02$ & $1034.79$ & $1034.75$ \\
SH0ES ($M_b$) & $8.91$ & $6.02$ & $1.48$ \\
BBN & $1.61$ & $2.15$ & $3.34$ \\
\hline
$\chi^2_{\rm tot}$ & $3825.35$ & $3822.11$ & $3814.65$ \\
$\Delta$AIC & $0.00$ & $0.76$ & $-4.71$ \\
\hline\hline
\end{tabular}
\caption{Full results from analyses with BBN performed in this work.  All analyses use the same data combination of Planck 2018, TT,TE,EE+lowE and lensing~\cite{Planck2018}; BAO from BOSS DR12~\cite{Alam_2017}, 6dF~\cite{Beutler_2011}, and MGS~\cite{Ross_2015}; and cosmic distance ladder measurements from Pantheon~\cite{Scolnic_2018} and SH0ES~\cite{Riess_2021}, as well as the BBN likelihood described in text.}
\label{tab:results_numbers}
\end{table}

\bibliographystyle{jcap}
\bibliography{references}

\end{document}